\documentclass[
               biblatex,     
               ]{jacow}

\usepackage[english]{babel}
\usepackage{graphicx}
\usepackage{biblatex}

\begin{document}

\title{DIAGNOSING GHOST BUNCHES WITH THE UPSTREAM EXTINCTION MONITOR IN THE \NoCaseChange{Mu2e} EXPERIMENT \thanks{This work was supported by the U.S. Department of Energy, Office of Science, Office of High Energy Physics under Award Number DE-SC0009999.} \thanks{Fermilab Report Number: FERMILAB-CONF-26-0327-PPD}}

\author{R.~Hensley\thanks{rshensley@ucdavis.edu}, E.~Prebys, University of California, Davis, Davis, United States\\
A.~Gaponenko, Fermi National Accelerator Laboratory, Batavia, United States}

\maketitle


\begin{abstract}
The Mu2e experiment has a stringent requirement for extinction of the pulsed proton beam, referring to the elimination of particles between proton bunches to a relative level of $10^{-10}$, which means a single out-of-time particle in the inter-pulse gaps for every 250 complete proton pulses.
As the construction of the Mu2e experiment nears completion, it is crucially important to make an early measurement of the beam extinction in its current condition.
Hence the upstream extinction monitor was constructed and operated to probe for problems in the proton pulse structure or a higher than expected incidence rate of out-of-time particles.

The analysis in this work comes from data taken in March 2026.
The long data run showed a significant presence of out-of-time particles from ghost bunches in the Delivery Ring approximately 388 ns after the centers of the main proton pulses.
These are hypothesized to be the result of a combination of a RF frequency mismatch, particle space charge, and machine impedance during the rebunching sequence in the Recycler Ring, which can lead to particles leaking into adjacent buckets, but further studies and simulations are needed to confirm this.

\end{abstract}

\section{Introduction and Monitor}

The Muon-to-Electron Conversion Experiment (Mu2e) at Fermilab \cite{mu2e-2014-tdr} searches for the neutrinoless conversion of a muon into an electron in the field of a nucleus, which would be a clear sign of physics beyond the Standard Model.
The experiment uses a pulsed proton beam to produce muons, and has a stringent requirement for beam extinction, which refers to the elimination of out-of-time particles between the proton bunches.
Basic extinction is defined as the ratio of out-of-time particles to in-time particles, and must be at a level of $10^{-10}$, which means a single out-of-time particle in the inter-pulse gaps for every 250 complete proton pulses.

\begin{figure}[!t]
    \centering
    \includegraphics[width=1.0\linewidth]{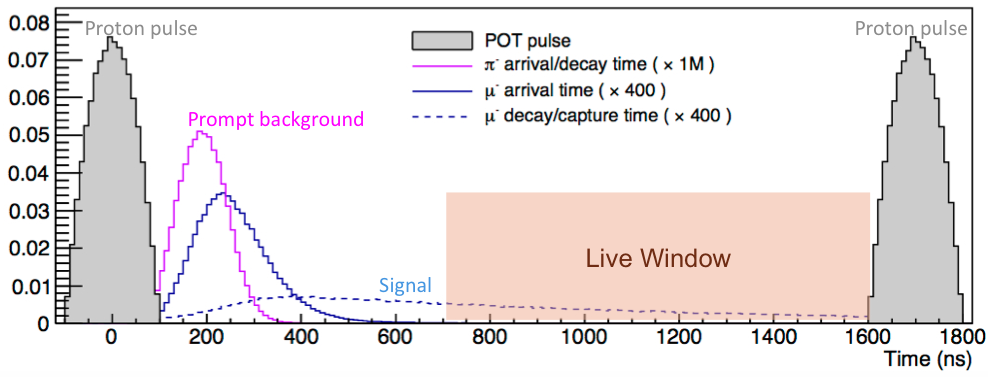}
    \caption{Mu2e proton-pulse timing and the live search window between pulses.}
    \label{fig:mu2e-timing-window}
\end{figure}

\begin{figure}[!t]
    \centering
    \includegraphics[width=0.8\linewidth]{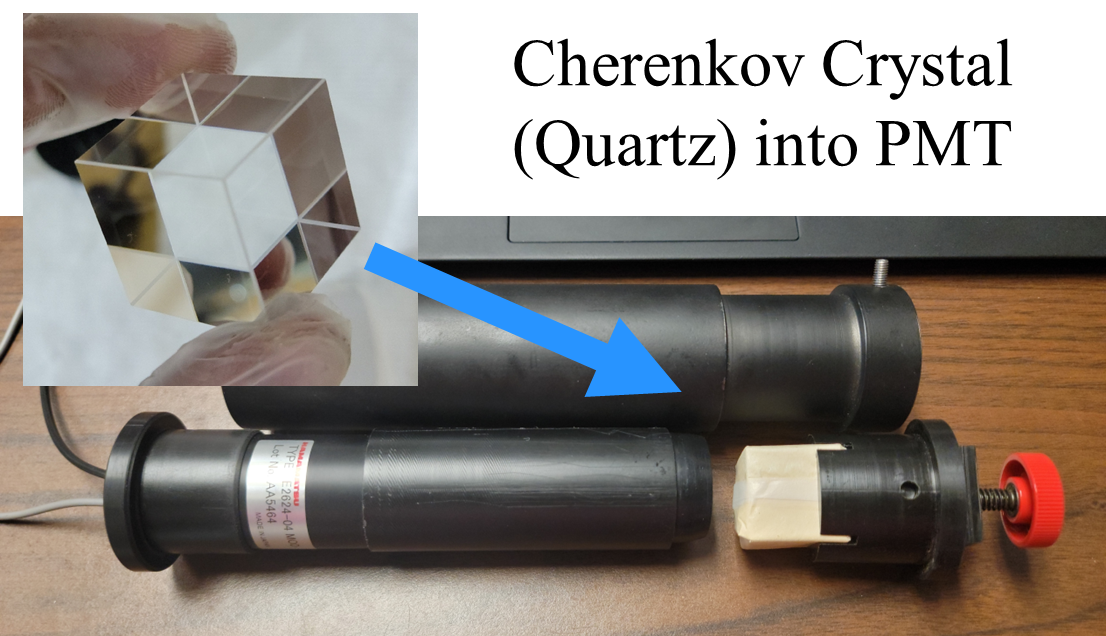}
    \caption{A quartz Cherenkov radiator with PMT assembly used in the Upstream Extinction Monitor.}
    \label{fig:mu2e-uem-detector}
\end{figure}

The beam extinction is crucial for the experiment, as out-of-time particles can produce background events that mimic the signal of muon-to-electron conversion, and therefore must be minimized to achieve the desired sensitivity.
The Mu2e proton-pulse timing and a representative Upstream Extinction Monitor detector element are shown in Figs.~\ref{fig:mu2e-timing-window} and~\ref{fig:mu2e-uem-detector}.

\subsection{Beam Extinction}

Proton delivery to the Delivery Ring starts with RF manipulations in the Recycler Ring.
The 53 MHz RF system quickly turns off, after which the 2.5 MHz RF system adiabatically ramps up, reshaping batches from the Booster Ring in preparation for transfers of individual bunches of protons into the Delivery Ring (note the frequency mismatch between the 53 MHz buckets and the 2.5 MHz buckets).
Once one bunch is orbiting the Delivery Ring, resonant extraction starts wherein one small pulse of protons is transferred into the M4 Beamline each time the proton bunch completes a revolution.
Resonant extraction completes after approximately 25,440 revolutions of the bunch around the Delivery Ring.
Given the \SI{1695}{ns} Delivery Ring revolution period, this establishes the basic time structure for the pulsed proton beam at an estimated preliminary extinction level of $10^{-4}$.

The remaining portion of the $10^{-10}$ extinction requirement is achieved via the operation of a system known collectively as the AC Dipole, a series of collimators and dipole magnets synchronized in phase to the transfers of the proton pulses.
The dipole magnets give transverse kicks to any out-of-time beam while allowing in-time particles to pass through unaffected.

\begin{figure*}[!t]
    \centering
    \includegraphics[width=0.8\textwidth]{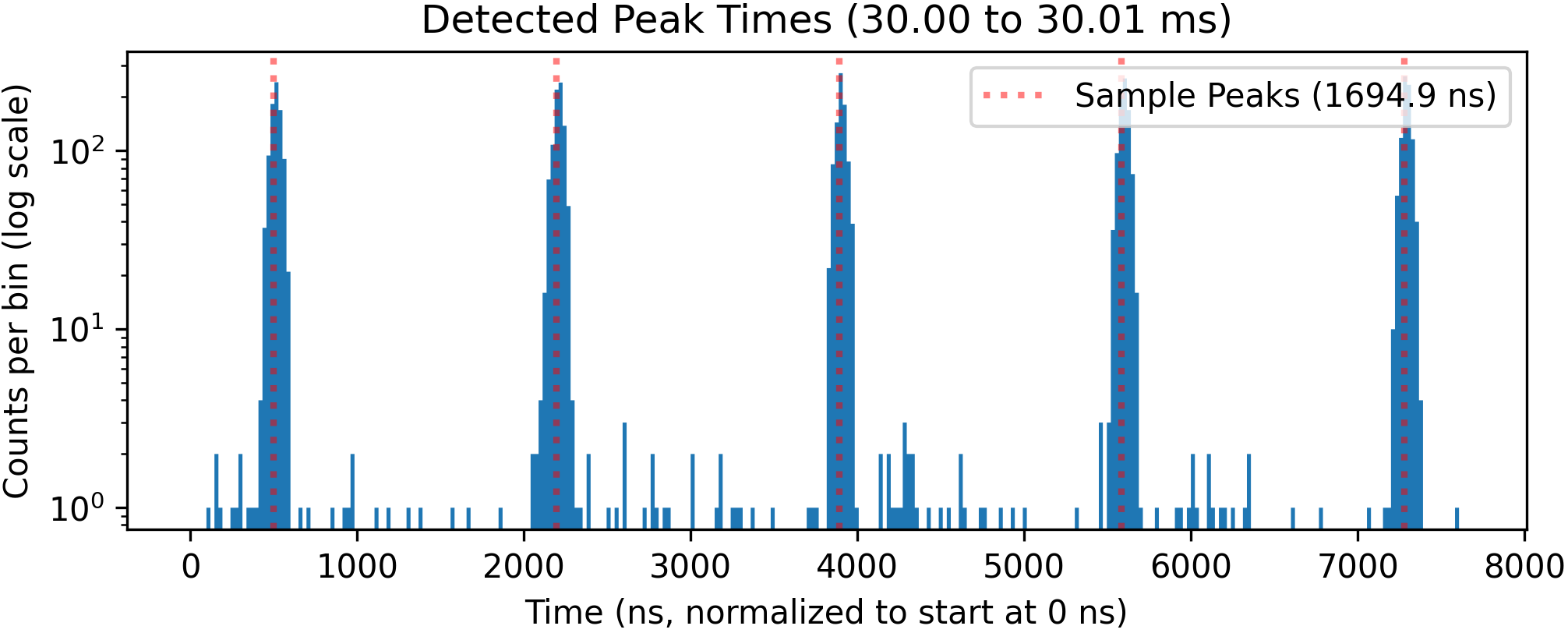}
    \caption{Zoomed time histogram of UEM peak data. The red dashed lines are manually overlaid to show the expected periodicity using the measured Delivery Ring revolution period.}
    \label{fig:wide-time-profile}
\end{figure*}

\subsection{Upstream Extinction Monitor Setup}

Previous work has been done to install and commission an Upstream Extinction Monitor\cite{hensley:ipac2025-thpm091,hensley:napac2025-mop037} in the M4 beamline.
It consists of a series of three photomultiplier tubes (PMTs) fixed to quartz Cherenkov radiators, and arranged in a line pointing at a vacuum window in the beam that serves as a scattering source.
With this, a parasitic measurement of the beam time profile can be made.
The goal is to obtain an early measurement of the longitudinal beam structure of the pulsed proton beam and provide an estimate to the extinction so problems with the beam can be remedied before the scheduled start of data taking in Spring 2028.

\section{Data and Timing Reconstruction}

\subsection{Run Sample}

In March 2026, data was accumulated from 914 Delivery Ring injections ($23.3 \times 10^6$ resonantly extracted proton pulses).
Data from the PMTs is sent to an ADC (AD9234) and FPGA (Kintex-7) housed in a Vadatech $\mu$TCA crate (AMC502, FMC228), where full peakfinding analysis occurs inside the FPGA, after which peak data is sent to a local computer for offline analysis.

Once offline, a fast Fourier transform (FFT) of the data is taken to estimate the revolution period of the Delivery Ring and spacing between proton pulses.
Fine adjustments to the period are necessary in order to be able to align $\sim$~$\SI{43.1}{ms}$ of data per event into a single \SI{1695}{ns} window~\cite{hensley:napac2025-mop037}.
A representative zoomed time window is shown in Fig.~\ref{fig:wide-time-profile}.

\section{Ghost-Bunch Observation}

\subsection{Spill-Time Structure}

\begin{figure}[!b]
    \centering
    \includegraphics[width=\linewidth]{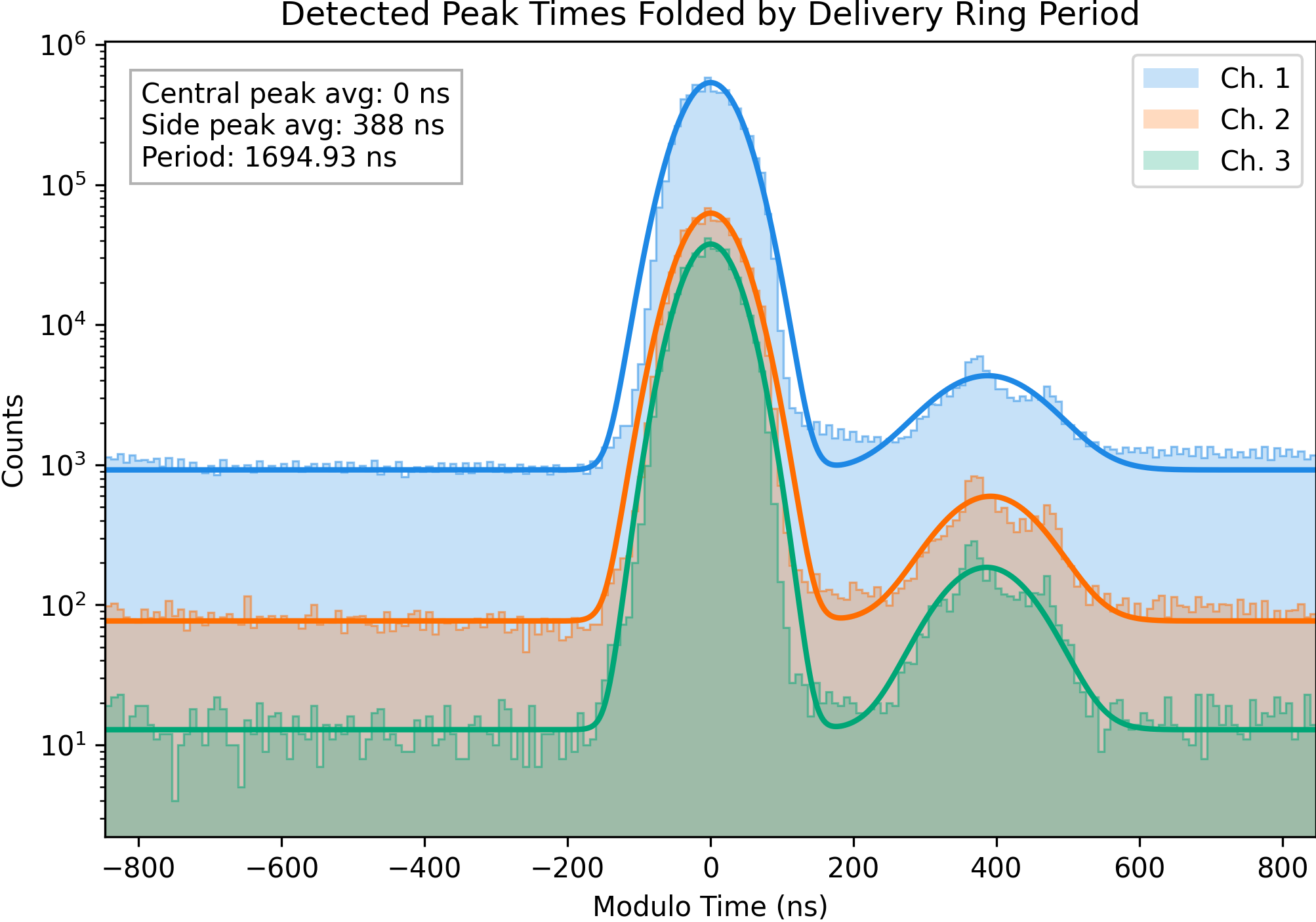}
    \caption{Period-modulated histogram of detected peaks. The main proton bunch is centered near zero, while the side peak appears in the adjacent timing bucket.}
    \label{fig:folded-profile}
\end{figure}

\begin{figure}[!t]
    \centering
    \includegraphics[width=0.9\linewidth]{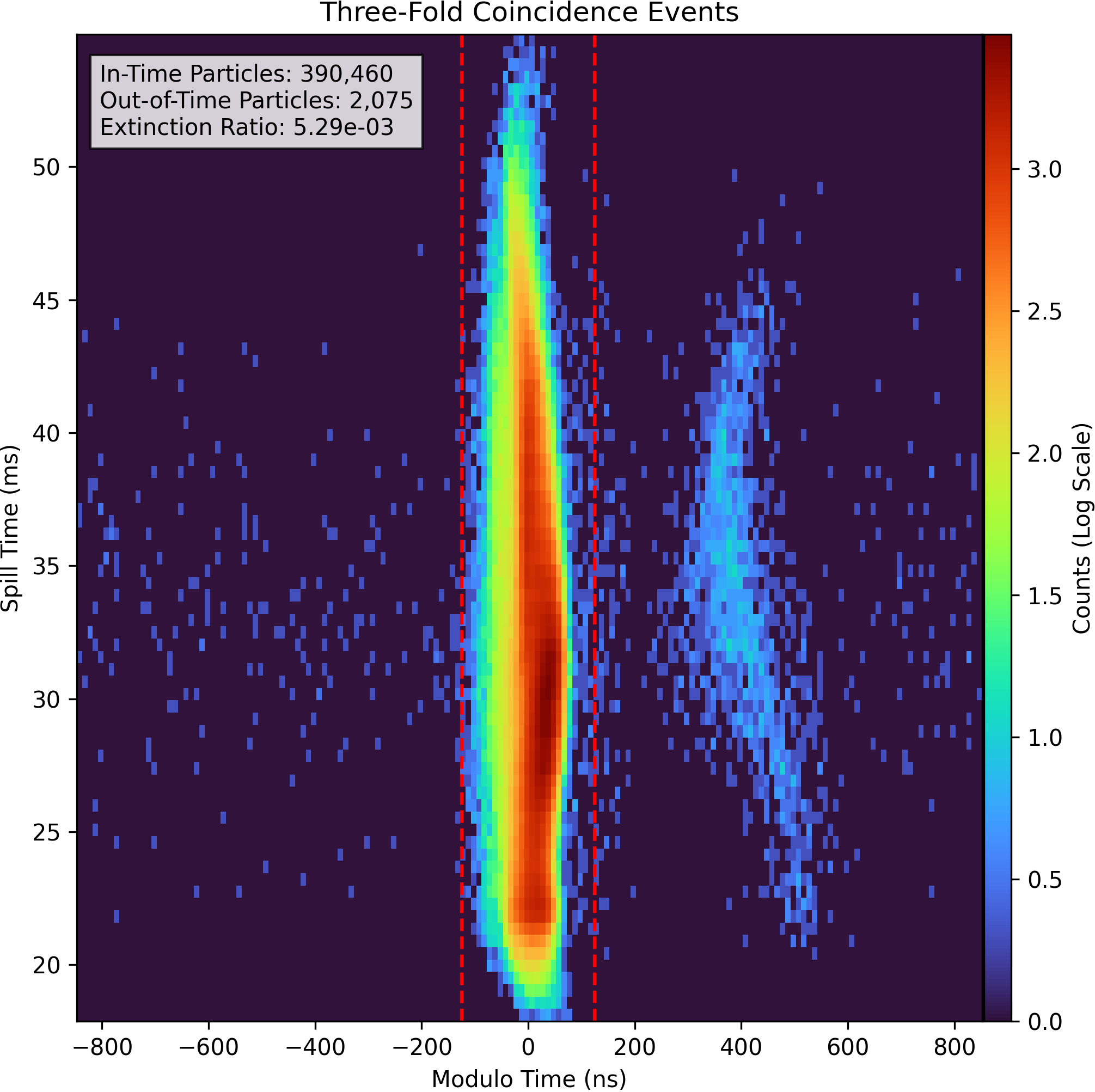}
    \caption{Three-fold coincidence events plotted as spill time versus modulo time. The main proton bunch appears near zero modulo time, while the ghost-bunch population appears in the adjacent bucket.}
    \label{fig:three-fold-coincidence}
\end{figure}

After folding all individual proton bunches over a spill (nominally \SI{43.1}{ms} in length) on top of each other (modulating the detected time by the FFT-derived Delivery Ring period), the expected shape of a nominal proton bunch can be seen with an approximate width of $\pm$\SI{250}{ns} in the center of the window, as shown in Fig.~\ref{fig:folded-profile}.
Combining the three data channels into a three-fold coincidence mode gives Fig.~\ref{fig:three-fold-coincidence}.
If all were ideal, the ratio of the number of particles outside this center pulse to the number of particles in the center would be $10^{-10}$.
However, it can be seen that there is a very noticeable bump to the right of the main proton pulse.

These are approximately \SI{388}{ns} after the main proton pulse is resonantly extracted, and are thought to be additional particles extracted from the (should-be-empty) adjacent RF bucket in the Delivery Ring.
This was an unexpected discovery, and it was not visible before without a long data run.
It has come to be referred to as a ``ghost bunch'' in the Delivery Ring.

\subsection{Possible Origins and Checks}

The direct cause of this ghost bunch is not clear.
However, a hypothesis can be made based on prior modeling and measurements made in the Recycler Ring.
These showed the generation of ghost bunches from two of the eight Recycler Ring bunches as a combined result of RF frequency mismatches, particle space charge effects, and machine impedance~\cite{harrig:2024vsn}.

In the Mu2e configuration, two adjacent Booster batches are transferred into the Recycler Ring, each containing 81 bunches of protons that fill 1/7 of the Recycler Ring.
Upon the arrival of the second Booster batch, the \SI{53}{MHz} RF system in the Recycler Ring is quickly turned off (\SI{5}{ms}) and then a different \SI{2.5}{MHz} RF system is adiabatically ramped in (\SI{85}{ms}).
Given that 53 and 2.5 are not harmonically related, this frequency mismatch in the RF systems can cause some particles to leak into adjacent buckets, which could be the source of the ghost bunches observed in the Delivery Ring, along with the fundamental effects of the space charge and the machine impedance.

\begin{figure}[!t]
    \centering
    \includegraphics[width=0.76\linewidth]{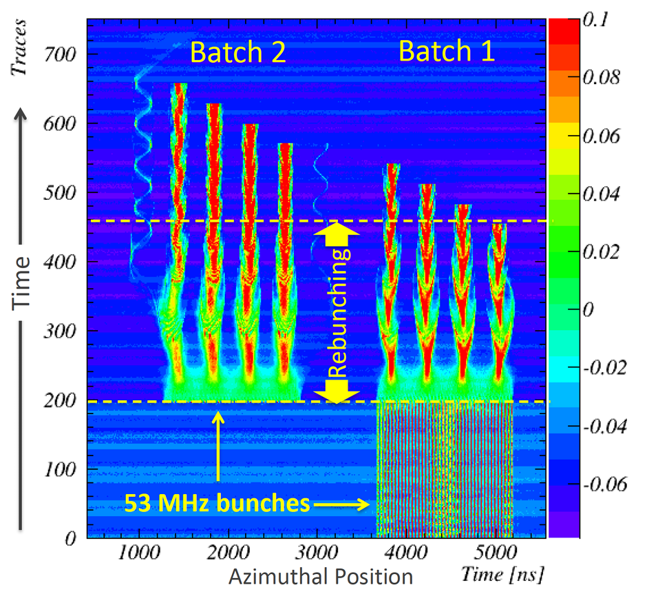}\\[-0.2em]
    \includegraphics[width=\linewidth]{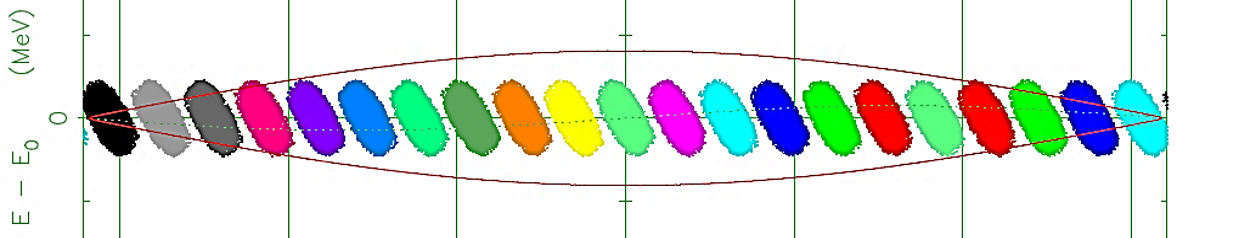}
    \caption{Recycler Ring measurements from previous rebunching studies and Beam Longitudinal Dynamics (BLonD) simulation framework output showing side populations adjacent to the primary bunch structure.}
    \label{fig:recycler-rebunching-evidence}
\end{figure}

This hypothesis is supported by modeling done in a simulation framework called Beam Longitudinal Dynamics (BLonD) of the rebunching sequence (shown in Fig.~\ref{fig:recycler-rebunching-evidence}), along with direct Recycler measurements, which show the rebunching of two Booster Ring batches into eight Recycler Ring bunches with ghost bunches appearing around Batch 2.
However, the Recycler extraction kicker could be eliminating these Recycler Ring ghost bunches, and other explanations are also possible, such as issues with the Recycler-to-Delivery Ring transfer given the \SI{2.5}{MHz} to \SI{2.36}{MHz} frequency mismatch there as well, and further studies and simulations are needed to confirm the true origin of the ghost bunches.
Potential solutions will be tested in the next beam testing period, starting with potentially extending the length of the Recycler Ring extraction kicker, or pre-firing an extra kicker.
Furthermore, simulation work is currently underway to extend BLonD simulations from the Recycler Ring into the Delivery Ring.

\section{Conclusion}

A March 2026 data run with the Upstream Extinction Monitor has revealed the presence of ghost bunches in the Delivery Ring, which are out-of-time particles that occur approximately 388 ns after the centers of the main proton pulses.
The background caused by the insufficiency in beam extinction from these ghost bunches would present a critical failure to the running of the Mu2e experiment.
Therefore, they must be understood and mitigated before the start of data taking in Spring 2028.

A leading hypothesis for the origin of these ghost bunches is a combination of a RF frequency mismatch during the rebunching sequence in the Recycler Ring, as well as fundamental particle space charge and machine impedance. Together, these effects can lead to particles leaking into adjacent buckets.
However, further studies and simulations are needed to confirm this hypothesis and rule out other potential explanations, such as issues with the Recycler-to-Delivery Ring transfer.
Follow-up beam studies will be conducted in the next beam testing period, starting with potentially extending the length of the Recycler Ring extraction kicker, and simulation work is currently underway to extend BLonD simulations from the Recycler Ring into the Delivery Ring.

\printbibliography

@techreport{mu2e-2014-tdr,
    author = "Bartoszek, L. and others",
    collaboration = "Mu2e",
    title = {{Mu2e Technical Design Report}},
    note = "arXiv:1501.05241 [physics.ins-det]",
    eprint = "1501.05241",
    archivePrefix = "arXiv:1501.05241 [physics.ins-det]",
    primaryClass = "physics.ins-det",
    reportNumber = "FERMILAB-TM-2594, FERMILAB-DESIGN-2014-01, Fermilab, Batavia, IL, USA",
    number="FERMILAB-TM-2594, FERMILAB-DESIGN-2014-01, Fermilab, Batavia, IL, USA",
    doi = {10.48550/arXiv.1501.05241},
    month = "10",
    year = "2014",
}

@phdthesis{harrig:2024vsn,
    author = {K. Harrig},
    title = {{Longitudinal Beam Structure Simulations with Impedance Studies and Ferrite Characterization for the {Mu2e} Experiment}},
    school = {University of California, Davis},
    year = {2024},
    type = {Ph.D. thesis},
    url = {https://escholarship.org/uc/item/0qt0390p},
    language = {english}
}

@inproceedings{hensley:ipac2025-thpm091,
%note ={presented at IPAC'25, Taipei, Taiwan, Jun. 2025, paper THPM091, to be published in the proceedings},
eventtitle={IPAC'25},
    author = {R. Hensley and others},
    title = {{Extinction Monitoring of Pulsed Proton Beams Using {FPGA}-Based Peak Detection}},
    booktitle = {Proc. IPAC'25},
    booktitle_alt = {Proc. 16th International Particle Accelerator Conference},
    pages = {2878-2881},
    eid = {THPM091},
    venue = {Taipei, Taiwan},
    series = {IPAC'25 - 16th International Particle Accelerator Conference},
    number = {16},
    publisher = {JACoW Publishing, Geneva, Switzerland},
    month = {06},
    year = {2025},
    issn = {2673-5490},
    isbn = {978-3-95450-248-6},
    doi = {10.18429/JACoW-IPAC25-THPM091},
    url = {https://indico.jacow.org/event/81/contributions/8345},
    language = {English},
    reportNumber = {value},
}

@inproceedings{hensley:napac2025-mop037,
    author = {R. Hensley and E. Prebys and A. Gaponenko},
    title = {{Calculating Beam Extinction in a Pulsed Proton Beam Using {FPGA}-Based Peak Detection}},
    booktitle = {Proc. NAPAC'25},
    pages = {127--130},
    paper = {MOP037},
    venue = {Sacramento, CA, USA, Aug. 2025},
    publisher = {JACoW Publishing, Geneva, Switzerland},
    doi = {10.18429/JACoW-NAPAC2025-MOP037},
    url = {https://proceedings.jacow.org/napac2025/pdf/MOP037.pdf},
    language = {english},
    reportNumber = {FERMILAB-CONF-25-0994-PPD},
}

\end{document}